\documentclass[12pt]{iopart}
\usepackage{iopams}
\usepackage{amssymb}
\usepackage{graphicx}
\usepackage{multirow}  
\begin{document}

\title[Super quantum probabilities and three-slit experiments]{Super quantum probabilities 
and three-slit experiments - Wright's pentagon state and the Popescu-Rohrlich box require third-order interference}

\author{Gerd Niestegge}
\address{Fraunhofer ESK, Hansastrasse 32, 80686 Muenchen, Germany}
\ead{gerd.niestegge@esk.fraunhofer.de}

\begin{abstract}
Quantum probabilities differ from classical ones in many ways, e.g., by violating
the well-known Bell and CHSH inequalities or another simple inequality due to R. Wright.
The latter one has recently regained attention because of its equivalence to a  
novel noncontextual inequality by Klyachko et al. 
On the other hand, quantum probabilities still obey many limitations 
which need not hold any more in more general probabilistic theories (super quantum probabilities).
Wright, Popescu and Rohrlich identified states which are included in such
theories, but impossible in quantum mechanics, and they showed this 
using its Hilbert space formalism.
Recently, Fritz et al. and Cabello detected 
that the impossibility of these states can be derived from 
very general principles (local orthogonality and global exclusive disjunction, respectively) 
without using Hilbert space techniques.
In the paper, an alternative derivation from rather different phyisical principles will be presented.
These are a reasonable calculus of conditional probability
(i.e., a model for the quantum measurement process)
and the absence of third-order interference.
The concept of third-order interference was introduced by Sorkin
who also recognized its impossibility in quantum mechanics.
\end{abstract}

\section{Introduction}
Quantum probabilities differ from classical ones in many ways, e.g., by violating  
the well-known Bell and CHSH inequalities \cite{ref-Bel, CHSH} or another simple inequality due to R. Wright \cite{Wri1978}.
The latter one has recently regained attention because of its equivalence to a  
novel noncontextual inequality by Klyachko et al. \cite{AH2013, BBCGL, CSW, Cab, Kly2008, LSW}.

On the other hand, quantum probabilities still obey many limitations 
which need not hold any more in more general probabilistic theories (super quantum probabilities).
The study of these limitations as well as of the differences from classical probabilities
is a current subject of research in the information theoretic and probabilistic foundations of quantum mechanics. 

A very general probabilistic theory is provided by the quantum logics \cite{BC} with a sufficiently rich state space, but
most of them do not allow for a reasonable calculus of conditional probability
and thus lack in a model for the quantum measurement process. 
However, those ones which entail a reasonable calculus of conditional probability yield a very fertile mathematical structure
\cite{Nie-AMP-2012}; they shall be called UCP quantum logics and provide the 
probabilistic framework in this paper. 

Wright \cite{Wri1978} as well as Popescu and Rohrlich \cite{PR} identified 
limitations of quantum mechanics in the form of states 
which are included in some general probabilistic 
theories, but impossible in quantum mechanics, and they showed this 
using its Hilbert space formalism.
Recently, Fritz et al. \cite{Fritz+} and Cabello \cite{Cab} detected 
that the impossibility of these states can be derived from 
very general principles (local orthogonality and global exclusive disjunction, respectively) 
without using Hilbert space techniques. An alternative derivation from 
a rather different physical principle will be presented in this paper.
This is the absence of third-order interference. 
The concept of third-order interference was introduced by Sorkin
who also recognized that third-order interference is ruled out by quantum mechanics \cite{Sor1994}.
His concept was 
adapted to conditional probabilities by Barnum, Emerson and Ududec \cite{BEU}.

Sections 2 and 3 briefly sketch the results from 
\cite{Nie-AMP-2012} concerning the quantum logics 
with a reasonable calculus of conditional probability and 
Sorkin's concept of third-order interference. 
In section 4, Wright's pentagon state is considered
and it is shown that it requires
third-order interference. This result is then applied 
to the Popesku-Rohrlich box in section 5.

\section{The calculus of conditional probability}

In quantum mechanics, the measurable quantities of a physical 
system are re\-presented by observables. Most simple are those 
observables where only the two discrete values 0 and 1 are 
possible as measurement outcome; these observables are called \textit{events} (or \textit{propositions})
and are elements of a mathematical structure called \textit{quantum logic} \cite{BC}.
Quantum mechanics uses a very special type of quantum logic; 
it consists of the self-adjoint projection operators on a 
Hilbert space or, more generally, in a von Neumann algebra.

An abstractly defined quantum logic $E$ contains two specific elements $0$ and $\mathbb{I}$ 
and possesses an \textit{orthogonality relation} $\bot$,
an \textit{orthocomplementation} $E\ni e \rightarrow e' \in E$
and a \textit{partial sum operation} + which is defined 
only for orthogonal events ($e + e' = \mathbb{I}$).
The interpretation of this mathematical terminology is as follows: 
orthogonal events are exclusive, $e'$ is the negation of $e$, and
$e + f$ is the disjunction of the two exclusive events $e$ and $f$.

The states on a quantum logic are the analogue of the 
probability measures in classical probability theory, and 
conditional probabilities can be defined similar to 
their classical prototype \cite{Nie-AMP-2012}. 
A \textit{state} $\mu$ allocates the probability $ \mu(f) \in [0,1]$ to each 
event $f$, 
is additive for orthogonal events, and $\mu(\mathbb{I})=1$.
The \textit{conditional 
probability} of an event $f$ under another event $e$ is the 
updated probability for $f$ after the outcome of
a first measurement has been the event $e$; it is denoted 
by $ \mu(f \mid e) $. Mathematically, it is defined by the
conditions that the map $E \ni f \rightarrow \mu(f \mid e)$
is a state on $E$ and that the identity 
$ \mu(f \mid e) = \mu(f)/\mu(e)$ holds for all events 
$f \in E$ with $f \bot e'$. It must be assumed that $\mu(e) \neq 0$.

However, among the abstractly defined quantum logics, 
there are many where no states or no conditional 
probabilities exist, or where the conditional probabilities 
are ambiguous. Therefore, only those quantum logics where 
sufficiently many states and unique conditional 
probabilities exist can be considered a satisfying 
framework for general probabilistic theories.
They shall be called \textit{UCP quantum logics} in this paper
and have been studied in Ref. \cite{Nie-AMP-2012}. Some of the 
results will be needed in this paper and shall now be sketched briefly .

A UCP quantum logic $E$ generates an order-unit space $A$
(partially ordered real linear space with a specific norm; see \cite{Han})
and can be embedded in its unit interval $\left[0,\mathbb{I}\right]$
$:=$ $\left\{a \in A : 0 \leq a \leq \mathbb{I} \right\}$;
$\mathbb{I}$ becomes the order-unit, and
$e' = \mathbb{I} - e$ for $e \in E$. 
Each state $\mu$ on $E$ has a unique positive linear extension on $A$ 
which is again denoted by $\mu$ \cite{Nie-AMP-2012}.

For each event $e$ in $E$, there is a positive linear map 
$U_e : A \rightarrow A$ with the following properties:
$\mu(f \mid e)\ \mu(e) = \mu(U_e f)$ for all $f \in E$ and all states $ \mu $,
$\mu(U_e x) = \mu(x)$ for all $x \in A$ and any state $\mu$ with $\mu(e)=1$,
$U_e^{2} = U_e$, $e = U_e e = U_e \mathbb{I}$ and $0 = U_e f$ for $e \bot f$,
$f = U_e f$ for $e' \bot f$. 
In quantum mechanics, $U_e x$ is the operator product $exe$, 
which reveals the link to the quantum measurement process.

A linear map $T_e$ can now be defined for each $e \in E$ by 
$T_e(x) := \frac{1}{2} (x + U_e x - U_{e'}x)$, $x \in A$.
The properties of the maps $U_e$ above imply the following properties for these maps: 
$\mu(T_e x) = \mu(x)$ for all $x \in A$ and any state $\mu$ with $\mu(e)=1$,
$e = T_e e = T_e \mathbb{I}$, and $0 = T_e f$ for $e \bot f$. 
In quantum mechanics, $T_e x$ is the Jordan product $e \circ x = (ex + xe)/2$.

These maps $T_e$ ($e \in E$) and their properties will play a significant role
in the following sections. First, an interesting link between them 
and Sorkin's concept of third-order interference shall be considered.

\section{Sorkin's third-order interference} 

Sorkin \cite{Sor1994} introduced the following mathematical term $I_3$ 
for a triple of pairwise orthogonal events $ e_1 $, $ e_2 $ and $ e_3 $, 
a further event $f$ and a state $ \mu $:
$$
\begin{array}{rcl}
  I_3 & := & \mu(f \mid e_1 + e_2 +e_3) \, \mu(e_1 + e_2 + e_3) - \mu(f \mid e_1 + e_2) \, \mu(e_1 + e_2) \\
    &   & - \mu(f \mid e_1 + e_3) \, \mu(e_1 + e_3) - \mu(f \mid e_2 + e_3) \, \mu(e_2 + e_3) \\
    &   & +  \mu(f \mid e_1) \, \mu(e_1) + \mu(f \mid e_2) \, \mu(e_2) + \mu(f \mid e_3) \, \mu(e_3) \\
\end{array}
$$
He recognized that $ I_3 = 0 $ is universally valid in 
quantum mechanics. His original de\-fi\-ni\-tion refers to probability measures 
on `sets of histories'. Using conditional pro\-ba\-bi\-li\-ties,
$ I_3 $ gets the above shape, which was seen by Ududec, Barnum and 
Emerson \cite{BEU}. 

For the three-slit set-up considered by Sorkin, the identity  $I_3 = 0$ 
means that the interference pattern observed with three open slits is a simple 
combination of the patterns observed in the six different cases when only one 
or two of the three slits are open. 
Though Sorkin's theoretical discovery that this holds in quantum
mechanics goes back to 1994, experimental testing has begun only recently and
confirmed it to the accuracy achieved in the experiment \cite{ref-Sin}.  

The new type of interference which is present whenever $ I_3 \neq 0 $ 
holds is called \textit{third-order interference}. 
In Ref. \cite{Nie-AMP-2012}, it has been shown that a UCP quantum logic $E$
rules out third-order interference ($I_3 \equiv 0$)
if and only if the identity
$T_{e+f} x = T_e x + T_f x$
holds for all orthogonal event pairs $e$ and $f$ in $E$ and all $x$ in $A$.
Mathematically, this orthogonal additivity of $T_e$ in $e$ is a lot easier to handle than the 
equivalent identity $I_3 \equiv 0$ with the above definition of the rather intricate 
term $I_3$ which, however, may be more meaningful physically. The orthogonal additivity
of $T_e$ in its index $e$
will play a central role in the derivation of the result in the next section
and make it possible to basically mimic Wright's original proof for the 
impossibility of the Pentagon state in the more general setting.

\section{Wright's pentagon state}

Consider a state $\mu$, five events
$e_1$,...,$e_5$,
the sum of their probabilities $\sum \mu (e_k)$,  
and assume  
$e_1 \bot e_2$, $e_2 \bot e_3$, $e_3 \bot e_4$, $e_4 \bot e_5$ and $e_5 \bot e_1$.
With classical probabilities, orthogonal events are disjoint sets and 
the maximum for $\sum \mu (e_k)$ is 2 (Wright's inequality). 
This can easily been seen looking at Figure 1. The overlapping areas contribute twice to the sum
which thus reaches its maximum when the probability is concentrated on these areas 
and all other areas carry zero probability; this maximum is 2.
\newline

\hspace{2 cm}
\includegraphics[height=3cm]{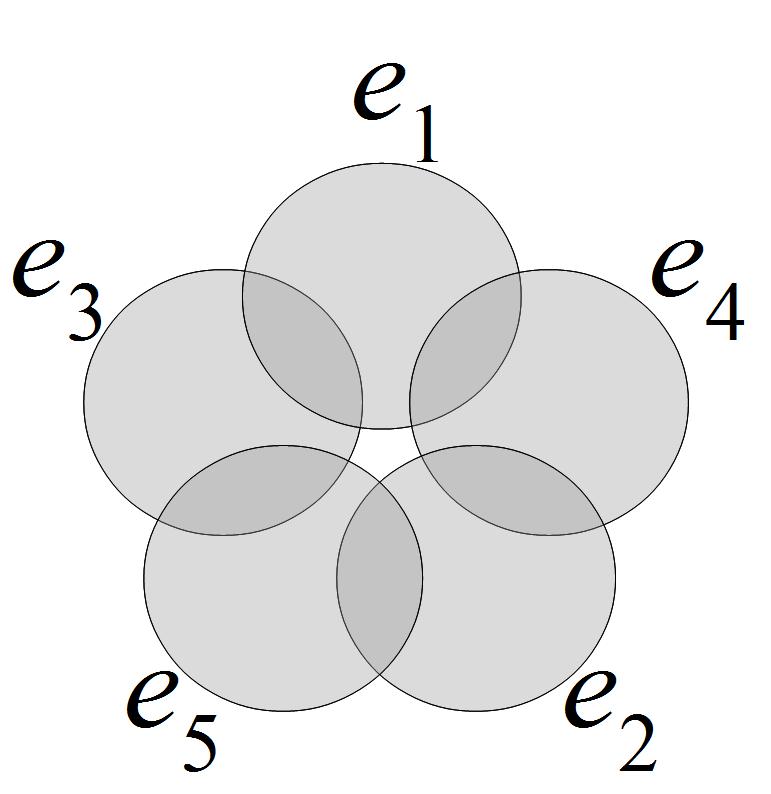}
\hspace{3,5 cm}
\includegraphics[height=3cm]{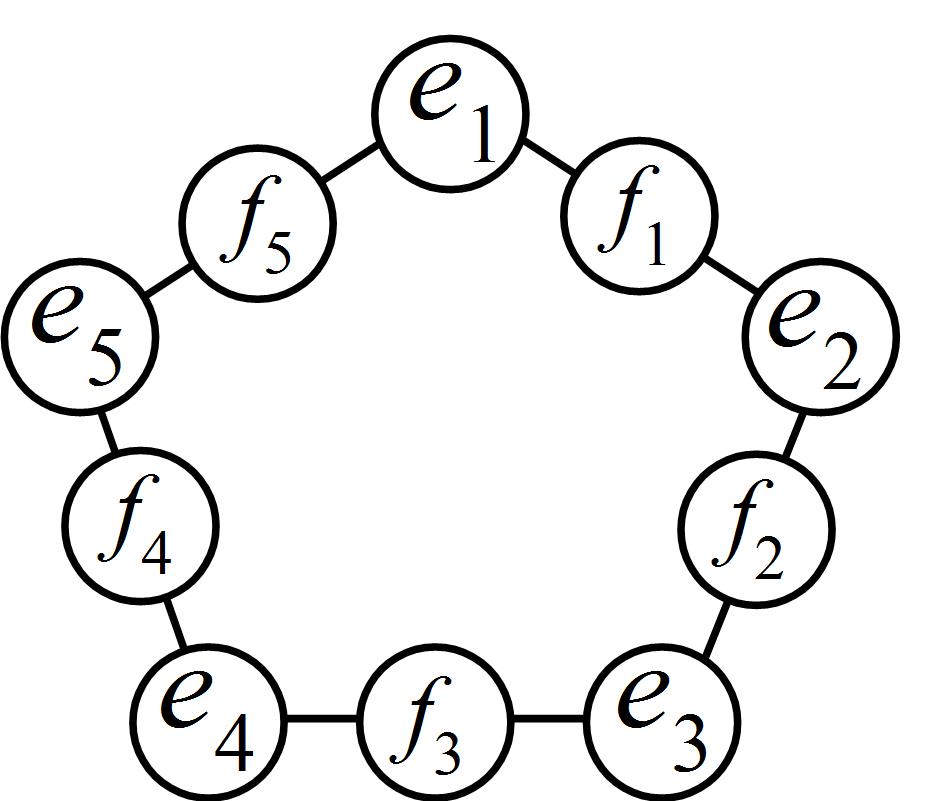}

\hspace{1 cm}
\textbf{Figure 1.} The classical case \hspace{1,4 cm} \textbf{Figure 2.} A quantum logic
\newline

In a quantum logic $E$, the situation is different. As an example, consider the quantum logic
with the Greechie diagram shown in Figure 2. Each one of the five straight lines 
represents a Boolean algebra $2^{3}$. A state on this quantum logic is defined by
$\mu (e_k) = 1/2$ and $\mu (f_k) = 0$ for $k=1,...,5$; this is Wright's pentagon state. 
Then $\sum \mu (e_k) = 5/2$.
On the other hand, $e_k \bot e_{k+1}$ for $k = 1,...,4$ and $e_5 \bot e_1$. 
Therefore $\mu (e_k) + \mu(e_{k+1}) \leq 1$ for $k = 1,...,4$, 
$\mu (e_5) + \mu(e_1) \leq 1$ and then $2 \sum \mu (e_k) \leq 5$. 
This proves that $5/2$ is the maximum for $\sum \mu (e_k)$ in general quantum logics 
and that this maximum can only be achieved if $1 = \mu (e_5) + \mu(e_1) = \mu (e_k) + \mu(e_{k+1})$,
$k = 1,...,4$. The only state satisfying this is the Pentagon state above,
since $\mu (e_1) = 1 - \mu (e_2) = \mu (e_3) = 1 - \mu (e_4) = \mu (e_5) = 1 - \mu (e_1)$ implies
$\mu (e_k) = 1/2$ for $k=1,...,5$. 

Applying the usual Hilbert space formalism, Wright showed that
the pentagon state is impossible in quantum mechanics \cite{Wri1978}. 
Now it will be seen that the
cal\-culus of conditional probability 
and the absence of third-order interference 
already suffice to rule out the pentagon state in a very broad probabilistic setting.

\textbf{Theorem:} $\sum\mu(e_k) \ ^{<}_{\neq} \ \frac{5}{2}$ 
if $e_1, ..., e_5$ lie in any UCP quantum logic $E$ with $I_3\equiv0$
and $ \mu $ is any state on $E$.

\emph{Proof}. Assume that $E$ is a UCP quantum logic with $I_3\equiv0$ and 
that $\sum\mu(e_k)=5/2$ for five events $e_1, ..., e_5$ in $E$ and a state $ \mu $ on $E$. 
This is possible only when
$\mu(e_k + e_{k+1}) = 1$ 
for $k=1,...,4$ as well as $\mu(e_1 + e_5) = 1$ (see above).
Using the general properties of the maps $T_e$ (section 2)
as well as their orthogonal additivity in $e$ implied by $I_3 \equiv 0$ (section 3),
it follows for any $x$ in $A$ and $k=1,...,4$ that 
$\mu(x) = \mu(T_{e_k + e_{k+1}}x) = \mu(T_{e_k}x) + \mu(T_{e_{k+1}}x)$
and $\mu(x) = \mu (T_{e_1 + e_5}x) = \mu (T_{e_1}x) + \mu(T_{e_5}x)$.
Subtracting each of these five identities from the next one results in 
$\mu(T_{e_{k-1}}x) = \mu(T_{e_{k+1}}x)$ for $k=2,3,4$, 
$\mu(T_{e_5}x) = \mu(T_{e_2}x)$  and $\mu(T_{e_4}x) = \mu(T_{e_1}x)$. 
Hence, $\mu(T_{e_1}x)$ = $\mu(T_{e_3}x)$ = $\mu(T_{5}x)$ = $\mu(T_{e_2}x)$ = $\mu(T_{e_4}x)$.
Finally, with $x=e_k$, $\mu(e_k) = \mu (T_{e_k}e_k) = \mu (T_{e_{k+1}}e_k) = 0 $ for $k=1,...,4$
and $\mu(e_5) = \mu (T_{e_5}e_5) = \mu (T_{e_1}e_5) = 0$,
which is a contradiction to $\sum\mu(e_k)=5/2$. $\Box$

For the study of contextuality, instead of $\sum\mu(e_k)$, Klyachko et al. \cite{Kly2008} consider 
$K := \mu (\sum^{4}_{k=1} x_k x_{k+1} + x_5 x_1)$
with the $\left\{+1,-1\right\}$-valued observables $x_k := 2e_k - 1$.
Then $x_k x_{k+1} = 1 - 2 e_k - 2 e_{k+1} $ and $K = 5 - 4\sum\mu(e_k)$ \cite{CSW}. 
It follows immediately that, under the assumptions of the theorem,
$K$ cannot reach its theoretical minimum $-5$.

In the following section, a further simple consequence of the theorem concerning   
nonlocality and the Popescu-Rohrlich box will be considered.

\section{The Popescu-Rohrlich box}

For the study of nonlocality, four $\left\{+1,-1\right\}$-valued observables 
$a_1, a_2, b_1, b_2$ are usually considered, where $a_1$ and
$a_2$ constitute a part of the system controlled by Alice 
and $b_1$ and $b_2$ a second 
part controlled by Bob.
Motivated from the spatial separation of the two parts, it is assumed that 
the joint probability distribution $p_{mn}$ of $a_m$ and $a_n$ exists for each $m,n=1,2$;
$p_{mn} (r,s)$ is the probability of the 
measurement outcomes $a_m = r$ and $b_n = s$ for $r=\pm1$, $s=\pm1$.
Relativistic causality requires the so-called
\textit{no-signaling principle}:
$p_{m1} (r,+1) + p_{m1} (r,-1) = p_{m2} (r,+1) + p_{m2} (r,-1)$ and
$p_{1n} (+1,s) + p_{1n} (-1,s) = p_{2n} (+1,s) + p_{2n} (-1,s)$
for $m,n=1,2$, $r,s=\pm1$.

\begin{center}
\begin{tabular}{c c|c c|c c|}
 & &  \multicolumn{2}{c|}{$a_1$} & \multicolumn{2}{c|}{$a_2$}  \\
 & & $+1$ & $-1$ & $+1$ & $-1$ \\
\hline
\multirow{2}{1mm}{$b_1$}
 & $+1$ & $0$ & $1/2$ & $0$ & $1/2$ \\
 & $-1$ & $1/2$ & $0$ & $1/2$ & $0$ \\
\hline
\multirow{2}{1mm}{$b_2$}
 & $+1$ & $0$ & $1/2$ & $1/2$ & $0$ \\
 & $-1$ & $1/2$ & $0$ & $0$ & $1/2$ \\
\hline
\multicolumn{6}{c}{\textbf{Table 1.} PR box}
\end{tabular}
\hspace{2 cm}
\begin{tabular}{c c|c c|c c|}
 & &  \multicolumn{2}{c|}{$a_1$} & \multicolumn{2}{c|}{$a_2$}  \\
 & & $+1$ & $-1$ & $+1$ & $-1$ \\
\hline
\multirow{2}{1mm}{$b_1$}
 & $+1$ &  & $e_5$ &  &  \\
 & $-1$ & $e_1$ & & $e_4$ & \\
\hline
\multirow{2}{1mm}{$b_2$}
 & $+1$ & & $e_2$ &  & \\
 & $-1$ &  &  &  & $e_3$ \\
\hline
\multicolumn{6}{c}{\textbf{Table 2.} $e_1,...,e_5$}
\end{tabular}
\end{center}

A measure for the statistical correlations
between the two systems are the expectation values $c_{mn}$
of the products $a_m b_n$ ($m,n = 1,2$). The maximum for
$ \left| c_{11} + c_{12} + c_{21} - c_{22} \right| $ 
is 2 in the classical case (CHSH inequality \cite{CHSH}) and
$2 \sqrt{2}$ in quantum mechanics (Tsirelson's bound \cite{Tsi}). 
The general algebraic maximum is 4 and can be reached 
without violating the no-signaling principle only by the so-called 
Popescu-Rohrlich boxes or, briefly, PR boxes \cite{PR}. 
One is shown in in Table 1.
The other seven PR boxes can be derived from  
Table 1 by exchanging $a_1$ with $a_2$, $b_1$ with $b_2$, or $+1$ with $-1$.

Again, the whole Hilbert space formalism of quantum mechanics 
is not necessary to rule out the PR boxes and the maximum 4 for 
$ \left| c_{11} + c_{12} + c_{21} - c_{22} \right| $, but the
calculus of conditional probability and the absence of third-order interference 
already suffice. This is an immediate consequence of the theorem in the last section
when applied to the events $e_1, ..., e_5$ as defined in 
Table 2 (see also \cite{Cab}). Note that it is assumed that two events 
occurring in the PR box scenario are orthogonal 
when they involve different values for the same observable.

A related stronger result has recently been presented in \cite{Nie-Tsi-2012} where
Tsirelson's bound is derived. However, it
requires some further mathematical assumptions.

\end{document}